\documentclass{epl}
\title{Beauty and Distance in the Stable Marriage Problem}
\author{G. Caldarelli\inst{1} \and A. Capocci\inst{2} }
\institute{
  \inst{1} INFM Unit\`a ROMA1 and Dip. di Fisica -
Universit\`a di Roma "La Sapienza", P.le A. Moro 2, 00185 Roma, Italy \\
  \inst{2} Institut de Physique Th\'eorique, Universit\'e de Fribourg, CH-1700
}
\pacs{05.20.-y}{Classical statistical mechanics}
\pacs{01.75.+m}{Science and society}
\pacs{02.50.Le}{Decision theory and game theory}

\begin{document}

\maketitle

\begin{abstract}
The stable marriage problem has been introduced in order to 
describe a complex system where individuals attempt to optimise their 
own satisfaction, subject to mutually conflicting constraints.
Due to the potential large applicability of such model to describe
all the situation where different objects has to be matched pairwise,
the statistical properties of this model have been extensively studied.
In this paper we present a generalization of this model, introduced in order
to take into account the presence of correlations in the lists and the
effects of distance when the player are supposed to be represented by a position
in space.
\end{abstract}

The problem of matching different players represents a 
well known problem in the area of operational research. 
A simple and sensible model one can build to face this 
situation is the stable marriage one.
In its traditional formulation \cite{GS62,Knu76,GusIrv89} 
one deals with two sets of 
$N$ players each of whom has an own list of the players
of the other set that he/she likes.
The first one in the list is the one he/she likes more, the last
one is the less liked.
It has already been fruitful to apply the ideas of the statistical 
physics in order to derive some properties of this model. 
In particular one would be tempted to study how ``to optimise''
the set of matchings in the system such that to maximise the 
general satisfaction.
Under this respect the stable marriage problem can be considered as
a prototype model for the social science and the economy 
where decision-makers are individuals or companies. They have their own  
selfish goals to optimise, which are often conflicting.
The stable marriage optimization problem does not 
require the total satisfaction to be the largest possible, 
but that the resulting state be stable against
the egoistic attempts of individuals to improve their
situation. This new concept of equilibrium is
familiar in game theory where anyone 
tries to maximise his utility at the same time. 
This is the concept of Nash equilibrium 
\cite{Nas50}, where a state is characterised by the 
stability with respect to the action of
any agent. In other words, a state is stable if any 
change in an agent's strategy is unfavorable for himself.
In this paper we use some of the results on the
statistical properties of the stable marriage problem
as obtained by Ref.\cite{Omero} \cite{Omero2}. We then generalise the model
by introducing amongst the different players lists a correlation
that should reduce the arbitrarity in the list compilation.
In other words our model refers to a more real world where the beauty
or the uglyness does exist and affects in similar way the players lists.
In this direction goes also the second generalization that we introduce
and study in this paper. We put the players on a lattice and we
consider explicitly the effect of distance in their matching.

The marriage problem describes a system where two sets of $N$ persons
have to be matched pairwise. 
In this formulation of the model all the players have to marry.
By suitably changing the language of the model, the same situation
can be applied to different contexts where two distinct sets 
have to be matched with the best satisfaction. 
Just to fix ideas we will use here the more intuitive language
of the man-woman matching, to explain the definition of the model.
Based on his judgment, each man establishes a wish-list of 
his desired women, in such a way that the top of the list is
the woman he likes more.  Women do the same for the men.
Following Ref.\cite{Omero} we assume also that 
each person's satisfaction depends on the rank
of the partner he/she gets to marry.
This allows us to regards the rank as a cost function to mimimise. 
If the top choice is attained, the cost is 
the least, the bottom choice has the highest cost.
We will consider in the following the case where the lists are 
random but not independently established. 

In order to set up the notation
let us look at the following example of three men 
and three women. The preference lists for all 
persons are shown below. We will use the symbol $m_i$
for the $i$-th man ($i=1,..,N$) and $w_i$ to index women, 
in such a way that a list ${\cal L}(m_i)$ of man $m_i$ can be described as: 

\begin{eqnarray}
{\cal L}(m_1) &=&\{w_1,w_2,w_3\} \nonumber \\
{\cal L}(m_2) &=&\{w_1,w_3,w_2\} \nonumber \\
{\cal L}(m_3) &=&\{w_2,w_3,w_1\} \nonumber \\ 
{\cal L}(w_1) &=&\{m_3,m_1,m_2\} \nonumber \\
{\cal L}(w_2) &=&\{m_1,m_2,m_3\} \nonumber \\
{\cal L}(w_3) &=&\{m_3,m_2,m_1\} 
\end{eqnarray}

If man $m_2$ marries woman $w_3$ (the second of his list) we assign
to the marriage a cost $2$ equal to the rank of his wife.
The minimum the cost the greater the satisfaction of the man.
It is convenient to introduce a representation of the lists in terms
of rankings: We define the matrices $F$  and $H$
for women and men respectively, such that the element $f(w_i,m_j)$
denotes the position of a man $m_j$ in the list of the woman $w_i$.
Equivalently $h(m_j,w_i)$ yields the rank of the woman $w_i$ in
the man $m_j$'s list. 
A realization of the preference lists is also called an instance 
and we indicate that as the set $\{m,w\}_i$ of the couples $i$ where 
$i=1,...,N$.
We call a {\em stable} matching ${\cal M} =\{m,w\}_i$ (where $i=1,..N$)
a state where one cannot find a man $m$ and a woman $w$ who are
not married but would {\em both}  prefer
to marry each other rather than staying with their respective
partners. Such a couple is called a {\em blocking pair}.
If no such pair exists, the matching is called stable.
One can calculate the energy per person, for 
women and men in a given  matching as
\begin{eqnarray}
\epsilon_F({\cal M})&=&\frac{1}{N}\sum_{i=1}^N f(w,m)_i \nonumber \\
\epsilon_H({\cal M})&=&\frac{1}{N}\sum_{i=1}^N h(m,w)_i
\end{eqnarray}
where $f(w,m)_i$ gives the rank of man $m$ for the woman $w$ in
the $i$-th couple, and 
where $h(m,w)_i$ gives the rank of woman $w$ for the man $m$ in
the $i$-th couple.
The energy $\epsilon({\cal M})=\epsilon_F({\cal M})+
\epsilon_H({\cal M})$ is the energy per couple.
Here and in the following the subscripts $F$ and $H$ stand for 
women and men, respectively.

To compute one stable state of the system it is usually used 
the classical Gale-Shapley (GS) algorithm  \cite{GS62} which
assigns the role of proposers to the elements of one set, the 
men say, and of judgers to the elements of the other. 
The {\em man-oriented} GS algorithm starts 
from a man $m$ making a proposal to the first woman $w$ on his 
list. If she accepts they get engaged, if she refuses $m$ goes 
on proposing to the next woman on his list. 
$w$ accepts a proposal when either she is not engaged or
she is engaged with a man $m'$ worse than the one 
proposing ($m$). In the latter case, $m'$ will proceed to the
woman following $w$ on his list. 
When all men have run through their lists proposing 
until all women are engaged, the algorithm
stops and the engagements result in marrings and 
matching thus reached is a stable matching.

As regards the previous example of a system with $N=3$ players, 
the man-oriented GS algorithm goes as follows:
man $m_1$ proposes to woman $w_1$ who accepts and they form the
pair $(m_1,w_1)$. Then $m_2$ proposes to $w_1$, but she 
refuses.
So man $m_2$ proposes to woman $w_3$ and they get married. 
Finally man $m_3$ happily marries woman $w_2$. This results
in the matching ${\cal M}_{H}=\{(m_1,w_1),(m_2,w_3),(m_3,w_2)\}$. 
If the GS algorithm is run by reversing the roles 
(woman-oriented) to yield the  woman-optimal
stable matching. In our example, this leads to 
${\cal M}_{F}=\{(m_1,w_1),(m_2,w_2),(m_3,w_3)\}$. 

The energies for men and women in these matchings are
$(\epsilon_H,\epsilon_F)=(4/3,7/3)$ and $(6/3,5/3)$ 
respectively for ${\cal M}_H$ and ${\cal M}_F$. 
As extensively studied in the ref.\cite{Omero}
who proposes is always better off than who judges. 
The man-oriented GS algorithm yields the man-optimal 
stable matching in the sense that no man 
can have a better partner in any other stable matching. 

All the above results hold in the case that the rank
any player assigns to the other sex players 
are randomly extracted from a uniform distribution.
A more realistic situation would apply if the players are
characterised  by what is commonly indicated by ``beauty''
that is some of the players have a finite probability to be 
high in rank in all the opposite sex players lists.
We tried to model the intrinsic beauty of players in such a 
way to generalize the ordinary stable marriage problem  that
can be derived now as a particular case.
For that reason in the new formulation we still have a random
contribution in the score a players receives from the opposite sex
players. Nevertheless a second contribution in the score formation
is given by a parameter weighting the intrinsic properties of the player.
In particular we assigned to each man/woman a parameter $I_{m}$ or $I_w$
respectively.
The score $S$ a woman receives in the opinion of a man is
given by 
\begin{equation}
S=\eta_{mw}+UI_w
\end{equation}
Where $\eta_m$ is a random number belonging to the 
interval $(0,1)$ reflecting the personal opinion
of man $m$ on woman $w$. The second term $I_w$ is again drawn 
from the interval $(0,1)$ and reflects the intrinsic
properties of woman $w$; it stays the same in all the scores 
for woman $w$ given by every man in the set.
$U$ is a parameter to weight the second contribution. 
At this point all the women have a score for the man $m$ that can
now form the list in descending order, that is the lower the score
the higher the  rank. In this way a small $I_w$ represents a 
nice woman for $U>>1$ (where the first contribution plays no role) 
conversely a value of $U=0$ corresponds to the standard stable marriage 
problem.
At this point we run the GS algorithm in order to detect 
the stable states.
As regards the statistical properties of the energy for such states we
start by considering the original case where the values of $f(w,m)_i$ and 
$h(m,w)_i$ differs each other.
This is the first plot present from above in Fig.1.
For this case no difference exists for the different players in the set 
and the average value of this rank over $1000$ different realisation
stays more or less the same.
This is not the case however when one rank the players in function of
their beauty such that the first ones are the more good-looking.
At this point the situation dramatically change and the satisfaction
now is dependent upon the intrinsic quality of the players.
This is evident in the second plot where $U$ is equal to $0.1$ and
even clearer in the last one where $U$ is equal to $1$.
It is interesting to note, however, that even if the more beautiful
players have by far a larger satisfaction in their matching with 
respect to the others, the general unsatisfaction in the system increases.
As a matter of fact, when the concept of ``most beautiful'' in the world
tends to be the same for everyone it becomes more and more difficult
to make more people happy.
However, the presence of beauty transforms in a fairer way the 
GS algorithm that now tends to give the same
results regardless the sex.
In all the plots of Fig.1 we called ``male'' those who propose and
``female'' those who judge. This means that ``male'' refers to both men in 
the man-oriented GS algorithm and to women in the woman-oriented
GS algorithm.
Another way to describe how correlations in lists affect the statistical
properties of the stable state, is to consider what is the probability 
$P(R)$ that a player will have a partner of rank $R$ in his/her list.
If in the case of uncorrelated lists it is easier to make more people happy,
this would reflect in a distribution of partner's rank peaked around the 
lowest values of rank. Conversely when beauty takes place one has to expect
a more spread distribution until the limit case where only beauty takes
place where one has to expect a uniform probability.
Indeed if $U\rightarrow \infty$ all the players have the same list.
They can be accomplished according to their own rank and since they are 
drawn from a uniform probability this results in this limit in a step 
function for the quantity $P(R)$.
Between these two limits, one can observe a transient behaviour
where $P(R)$ is a power law in $R$ as shown in Fig.2.

Another generalisation that we introduce in order to deal with
a realistic situation can be obtained by considering a spatial 
distribution of players and by adding an additional cost for all
the marriages that happen at a large distance.
In complete analogy to what  has been done for the beauty we will define
the score of woman $w$ for man $m$ as
\begin{equation}
S=\eta_{mw}+\alpha d(m,w)
\end{equation}
where $d(m,w)$ is the euclidean normalised distance between two players arranged on
a $d-dimensional$ lattice and $\alpha$ weights this contribution in the score
formation.
Since the first part of the score takes into account the  personal opinions
through a random number drawn from a uniform distribution between 
$0$ and $1$ we normalise the value of distance by dividing it for the
larger value of the distance possible in the system (i.e. in $d=1$ for
$N$ players we have a maximum of $N/2$).
In this way a value of $\alpha=0$ corresponds to the classical case, whilst
in the limit of $\alpha=0$ the dynamics is dominated by nearest neighbors
marriages.
In this case we did not found any difference from the classical case for
what regards the GS algorithm. Also in this case the man-oriented algorithm
and the woman-oriented brings to different stable states more convenient for
men and women respectively. 
The reason for that in our opinion is in the fact that different random lists
form different metastable states that GS can explore in the finding of stability.
This scenario is destroyed by the presence of correlations that tends
to form a common judgement for all the players (at least for those more beautiful 
that can choose).
When distance is taken into account, instead, this new effect does not produce 
a player that is most beautifulk than the other and that moves for everyone 
on the top of the lists. What happens iin this cases that for every player
one has a subset of the neighbours (and they are different for every player!)
that get some advantage in the matching.
To show this effect we plotted in Fig.3 how the probability function $P'(d)$ to
get married with a partner at a distance $d$ varies with $\alpha$.

We have generalised the classical stable marriage problem
to the more realistic situation when correlations are present in
the lists and when the effect of spatial distance between players
are taken into account.
The classical stable marriage
problem can be considered a prototype model for the game theory where
the concept of Nash equilibrium plays the central role. 
The game-theoretical definition of stability  
result in the formation of stable states that are not the globally best solution.
We show that the presence of correlation removes the difference in the stable
state that is produced when the utility of one kind of player is
maximised with respect to the opposite sex players.
The presence of correlation in the wish lists reduces the general satisfaction
in the system but distributes it in a fairer way.
A similar result is found when spatial distance between players is taken into account.

In conclusion our paper shows that in case of more realistic situations the 
framework of Nash equilibria for the game theory do not necessarily apply.
This result in a more conventional search for stable state in the problems 
of matching.

This work was supported by the European 
Network contract FMRXCT980183.

\begin{figure}
\onefigure{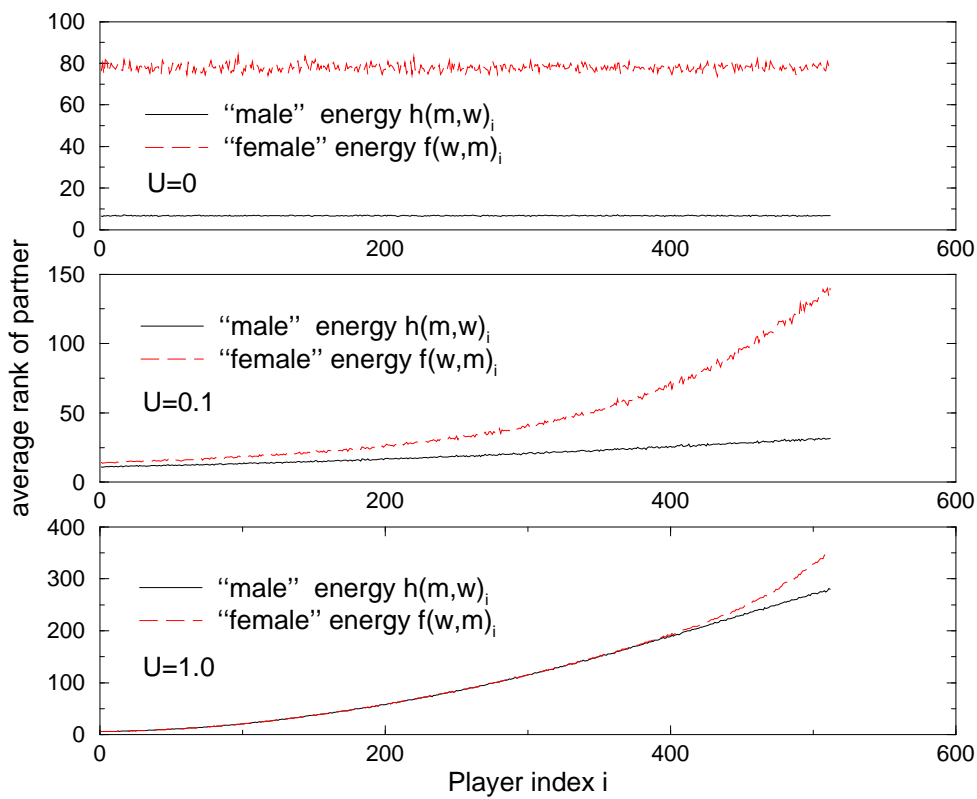}
\caption{Energy of men $(\epsilon_H)$ and women $(\epsilon_F)$
for all the players ($N=512$) averaged over $1000$ of stable 
matchings. above in the first plot we present the results corresponding to
the usual case of uncorrelated lists.
Proceeding below we present the same quantities as the parameter $U$ increases.}
\label{Fig.1}
\end{figure}

\begin{figure}
\onefigure{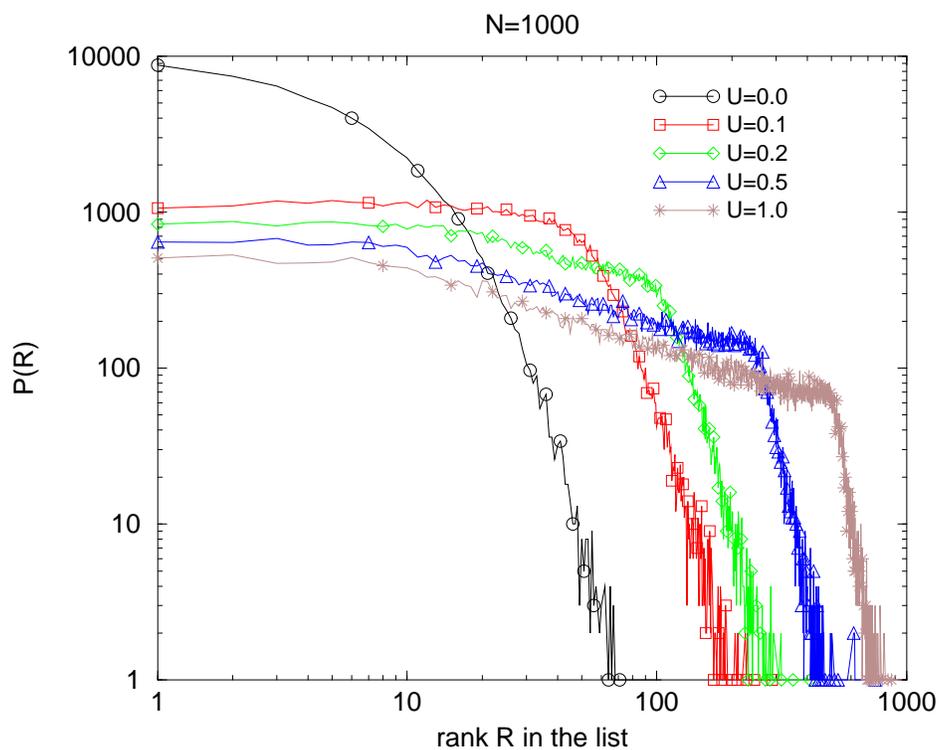}
\caption{
Density function $P(R)$ giving the probability that a male $m$ is married with 
a woman $w$ that is ranked $R$ in his list.
}
\label{fig2}
\end{figure}

\begin{figure}
\onefigure{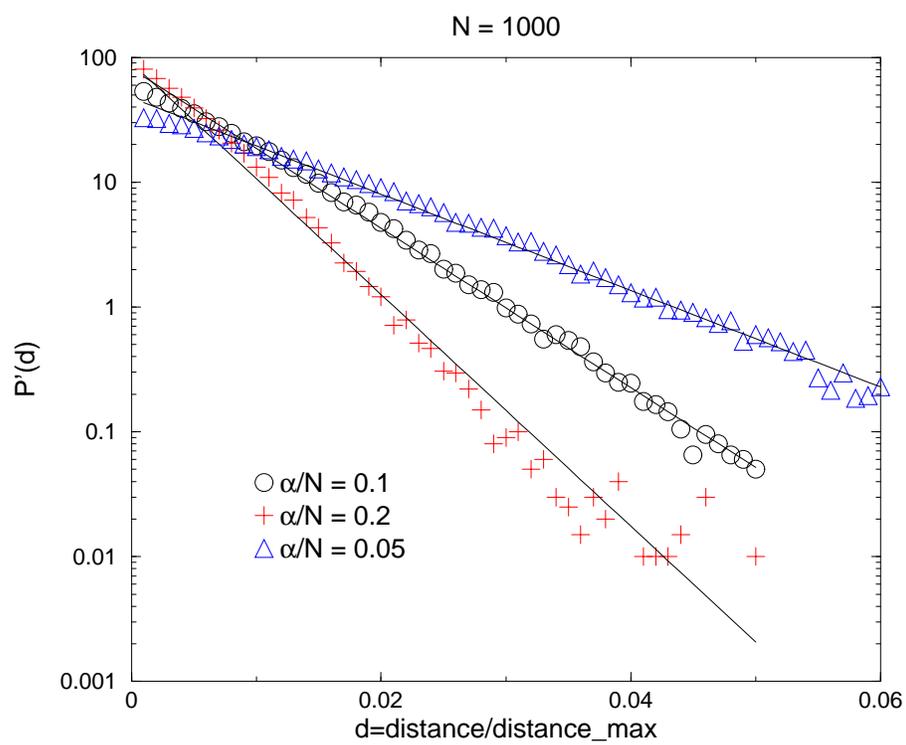}
\caption{
Density function P'(d) giving the probability to have a marriage where 
male and female are divided by a normalised distance $d$.
}
\label{fig3}
\end{figure}

\end{document}